# Bayesian Vote Manipulation: Optimal Strategies and Impact on Welfare

**Tyler Lu**
Dept. of Computer Science
University of Toronto

**Pingzhong Tang**
Computer Science Dept.
Carnegie Mellon University

**Ariel D. Procaccia**
Computer Science Dept.
Carnegie Mellon University

**Craig Boutilier**
Dept. of Computer Science
University of Toronto

## Abstract

Most analyses of manipulation of voting schemes have adopted two assumptions that greatly diminish their practical import. First, it is usually assumed that the manipulators have full knowledge of the votes of the nonmanipulating agents. Second, analysis tends to focus on the probability of manipulation rather than its impact on the social choice objective (e.g., social welfare). We relax both of these assumptions by analyzing *optimal Bayesian manipulation strategies* when the manipulators have only partial probabilistic information about nonmanipulator votes, and assessing the *expected loss in social welfare* (in the broad sense of the term). We present a general optimization framework for the derivation of optimal manipulation strategies given *arbitrary* voting rules and distributions over preferences. We theoretically and empirically analyze the optimal manipulability of some popular voting rules using distributions and real data sets that go well beyond the common, but unrealistic, impartial culture assumption. We also shed light on the stark difference between the loss in social welfare and the probability of manipulation by showing that even when manipulation is likely, impact to social welfare is slight (and often negligible).

## 1 INTRODUCTION

The use of voting rules to aggregate preferences has become a topic of intense study, and one of great importance in ranking, recommender systems, resource allocation, and other applications of social choice to computational systems. One of the most challenging topics in computational social choice is the study of *manipulation*: given a voting rule, there is usually some set of voter preferences such that one or more voters can obtain a more desirable outcome by misreporting their preferences. Indeed, except under very stringent assumptions, voting rules that are not manipulable do not exist [18, 33]. An important line of research, initiated by Bartholdi *et al.* [2, 3], shows that it can be computationally difficult to manipulate certain rules. However, since worst-case complexity is not itself a significant barrier to manipulation, recent attention has focused on theoretical and empirical demonstration that manipulation is often, or on average, tractable for certain stylized preference distributions [11, 31, 16, 35].

While our understanding of manipulation has improved immensely, some significant deficiencies remain in the state of the art. First, analysis of manipulation is often confined to cases in which the manipulators have complete knowledge of the preferences of sincere voters. While there are reasons for this, such analyses offer too pessimistic a picture of manipulability, since manipulators rarely have access to such information.

The main contribution of our work is a framework for analyzing *optimal Bayesian manipulation strategies* under realistic knowledge assumptions, namely, manipulators with *partial, probabilistic knowledge* of voter preferences. Since success of a manipulation is generally uncertain, the optimal strategy simply maximizes the odds of success. This depends critically on the voting rule, the number of voters, and the preference distribution—as a result, general analytical results are difficult to obtain. We instead present an empirical methodology that, assuming only the ability to sample vote profiles from the preference distribution, allows one to compute an optimal manipulation strategy. We illustrate this methodology on several preference distributions (including real-world data). We also derive sample complexity bounds that provide quality guarantees on the resulting strategies. This framework provides the ability to analyze the manipulability of most voting rules under any probabilistic knowledge assumptions—arbitrary priors, posteriors conditioned on available evidence, even the special case of complete knowledge—requiring only that the preference distribution be sampleable.

While the computational framework is general, we also analytically derive optimal manipulation strategies for the $k$-approval rule (and the Borda rule in a limited fashion) under standard *impartial (and anonymous) culture* assumptions. However, since impartial culture is rare in practice [32], these results are primarily of theoretical interest.

A second deficiency of current manipulation analyses pertains to the emphasis on success probability. We adopt a decision-theoretic perspective that provides a more nuanced picture of manipulability of various voting rules. Intuitively, the more preferred an alternative is to the sincere voters, the more likely it is that manipulators can succeed in causing the alternative to be selected. As such, probability of manipulation does not tell the whole story, since alternatives with higher success probability cause less societal dissatisfaction. To this end, we interpret the "score" associated with specific alternatives under a given voting rule as an explicit social choice objective—i.e., a *social welfare function* in the broad sense of the term—and propose analyzing rules in this light. We recognize that voting protocols are often applied in settings where maximizing (some form of) social welfare is not the main objective; but we argue below that this perspective is natural in many applications, and our methodology applies to measures of social welfare different from the "score" used by the voting rule itself.

Along these lines, we derive theoretical bounds on the impact of manipulation on social welfare for the special case of positional scoring rules. More importantly, while distribution-dependent analytical results are difficult to obtain, we can exploit our ability to identify optimal manipulation strategies: by sampling vote profiles from the distribution, computing the loss in welfare under the *optimal* manipulation strategy, and averaging the results, we can readily determine the expected impact on welfare for any specific distribution. Our empirical results show that manipulation for certain rules, under realistic knowledge assumptions, generally causes little damage.

## 2 BACKGROUND

We briefly review relevant background; see [17, 9] for further detail on (computational) social choice.

### 2.1 VOTING RULES

We assume a set of $n$ *voters* $N = \{1, \ldots, n\}$ and a set of $m$ *alternatives* $A = \{a_1, \ldots, a_m\}$. A voter $i$'s preferences are represented by a *ranking* or permutation $v_i$ over $A$. If $v_i(a)$ is the *rank* of $a$ in $v_i$, we say $i$ prefers $a_j$ to $a_k$ if $v_i(a_j) < v_i(a_k)$. We refer to $v_i$ as $i$'s *vote*. A *preference profile* is a collection of votes $\mathbf{v} = (v_1, \ldots, v_n)$. Let $\mathcal{V}$ be the set of all such profiles.

A *voting rule* $r : \mathcal{V} \to A$ selects a "winner" given a preference profile. Common voting rules include plurality, approval, Borda, single transferable vote (STV) and many others. We will use *positional scoring rules* frequently in the sequel. Let $\alpha = (\alpha_1, \ldots, \alpha_m)$ be a non-negative *positional scoring vector*, where $\alpha_i \geq \alpha_{i+1}$ for all $i \geq 1$. Intuitively, $a$ earns score $\alpha_i$ for every vote that places it in position $i$, and the alternative with the greatest total score wins.[1] More formally, let the $m \times m$ *positional summary matrix (PSM)* $\mathbf{X}$ for profile $\mathbf{v}$ have entry $X_{ij}$ equal to the number of votes that rank $a_i$ in position $j$. Then the *score* $s_\alpha(a_i; \mathbf{v})$ of $a_i$ is the $i$-th entry of $\mathbf{X}\alpha'$. Several important rules can be defined using positional scores: plurality by $(1, 0, \ldots, 0)$; $k$-approval by $(1, 1, \ldots, 0, 0)$, with $k$ 1s; and Borda by $(m-1, m-2, \ldots, 0)$.

Many voting rules—not just positional rules, but rules such as maximin, Copeland, Bucklin, Kemeny, and many others—*explicitly score all alternatives*, assigning a score $s(a, \mathbf{v})$ that defines some measure of the quality of alternative $a$ given a profile $\mathbf{v}$, then choosing the alternative with maximum score. This can be viewed as implicitly defining a *social choice objective* that is optimized by the voting rule, or more broadly as a *social welfare function*, that expresses some "societal utility" for each alternative.[2] While this interpretation may not be valid in all contexts, we will refer to $s(a, \mathbf{v})$ as the *social welfare* of $a$ under rule $r$ for voters with preferences $\mathbf{v}$, and write this as $SW(a, \mathbf{v})$ to emphasize this view. We discuss this further below.

### 2.2 MANIPULATION

By *manipulation* we refer to a coalition of one or more voters obtaining a more desirable outcome by misreporting their preferences. Indeed, except under very stringent conditions (e.g., single-peaked preferences), the classic Gibbard-Sattherwaite theorem shows that no scheme is immune to manipulation [18, 33]. In other words, there is some preference profile for which at least one voter can obtain a better outcome by misreporting. In the remainder of the paper, we focus on *constructive manipulation*, in which a coalition attempts to cause a single "preferred" candidate to win. We emphasize however that the general principles underlying our approach apply equally well to other forms of manipulation such as destructive manipulation or utility maximization (see below).

---

[1]In Sec. 3, we assume that tie-breaking works against the desired alternative of the manipulators.

[2]We we use the term "social welfare" in its broadest sense, referring to *any* means of ranking social outcomes. Specifically, we do *not* assume its more restricted definition, commonly used in mechanism design, as "sum of individual voter utilities."

Whether manipulation is a problem in practice depends on: how likely such manipulable preference profiles are; whether manipulators can detect the existence such a profile; and whether computing a suitable misreport is feasible. On this third point, the pioneering work of Bartholdi *et al.* [2, 3] demonstrated that, even given full knowledge of a profile, computing a suitable manipulation is computationally intractable for certain voting rules. This in turn led to the detailed computational analysis of many voting rules (e.g., the Borda rule [13, 4]). Of course, worst-case complexity results offer little comfort if "difficult profiles" are unlikely to arise in practice. Recent work suggests that common voting rules are in fact frequently manipulable, by studying heuristic algorithms that provide theoretical guarantees [31, 41, 39], identifying properties of voting rules that make them easy to manipulate in the typical case [11, 16, 40], and investigating (both theoretically and empirically) the relation between the number of manipulators and the probability of manipulation [30, 38, 35] (see [15] for an overview).

Analyses demonstrating ease of manipulation tend to suffer from two key drawbacks. First, they exclusively analyze manipulation assuming the manipulating coalition has full knowledge of the vote profile. While results showing that manipulation is *difficult* can be justified on these grounds, claiming *easiness* of manipulation has less practical import if the coalition is assumed to have unreasonable access to the preferences of sincere voters.[3] One exception considers manipulators who know only that the vote profile lies within some set [12], but unfortunately this work only analyzes the rather weak notion of *dominating manipulations*. Social choice research on manipulation under probabilistic knowledge is mostly negative in nature [23], or restricted to a single manipulator [1].

A second weakness of many analyses of probability of manipulation (which do assume complete information on the part of the manipulator) is their reliance on specific stylized models such as impartial culture (where every ranking in equally likely) [16, 40]. Much empirical work also considers very stylized distributions such as impartial culture, Polya's urn, and Condorcet (or Mallows) distributions. Some work does consider sub-sampling from real voting data, though with relatively small numbers of votes [36].

### 2.3 PROBABILISTIC RANKING MODELS

Probabilistic analysis of manipulation—including our Bayesian manipulation problem—requires some probabilistic model of voter preferences. By far the most common model in social choice is *impartial culture (IC)*, which assumes the preference of any voter is drawn from the uniform distribution over the set of permutations of alternatives [32]. A related model is the *impartial anonymous culture (IAC) model* in which each *voting situation* is equally likely [32].[4] Several other models (bipolar, urn, etc.) are considered in both theoretical and empirical social choice research.

Probabilistic models of rankings are widely considered in statistics, econometrics and machine learning as well, including models such as Mallows $\phi$-model, Plackett-Luce, and mixtures thereof [25]. We use the *Mallows $\phi$-model* [24] in Sec. 6, which is parameterized by a reference ranking $\sigma$ and a dispersion $\phi \in (0,1]$, with $P(r) = \frac{1}{Z}\phi^{d(r,\sigma)}$, where $r$ is any ranking, $d$ is Kendall's $\tau$-distance, and $Z$ is a normalizing constant. When $\phi = 1$, this model is exactly the impartial culture model studied widely in social choice—as such it offers considerable modelling flexibility. However, mixtures of Mallows models offer even greater flexibility, allowing (with enough mixture components) accurate modelling of any distribution over preferences. As a consequence, Mallows models, and mixtures thereof, have attracted considerable attention in the machine learning community [27, 8, 20, 26, 21]. We investigate these models empirically below.

## 3 OPTIMAL BAYESIAN MANIPULATION

We now consider how a manipulating coalition should act given *probabilistic* knowledge of the preferences of the sincere voters. We first formally define our setting, then present several analytical results. Finally, we present a general, sample-based optimization framework for computing optimal manipulation strategies and provide sample complexity results for positional scoring rules and $k$-approval.

### 3.1 THE MODEL

We make the standard assumption that voters are partitioned into $n$ sincere voters, who provide their true rankings to a voting mechanism or rule $r$, and a coalition of $c$ manipulators. We assume the manipulators have a *desired alternative* $d \in A$, and w.l.o.g. we assume $A = \{a_1, \ldots, a_{m-1}, d\}$. We make no assumptions about the manipulators' specific preferences, only that they desire to cast their votes so as to maximize the probability of $d$ winning under $r$. A vote profile can be partitioned as $\mathbf{v} = (\mathbf{v}_n, \mathbf{v}_c)$, where $\mathbf{v}_n$ reflects the true preferences of the $n$ sincere voters and $\mathbf{v}_c$ the *reported* preferences of the $c$ manipulators.

---

[3]This is implicit in [10], which shows that hardness of full-information manipulation implies hardness under probabilistic information.

[4]A voting situation simply counts the *number* of voters who hold each possible ranking of the alternatives.

In contrast to most models, we assume the coalition has only *probabilistic knowledge* of sincere voter preference: a distribution $P$ reflects these *beliefs*, where $P(\mathbf{v}_n)$ is the coalition's degree of belief that the sincere voters will report $\mathbf{v}_n$. We refer to the problem facing the coalition as a *Bayesian manipulation problem*. Manipulator beliefs can take any form: a simple prior based on a standard preference distributions; a mixture model reflecting beliefs about different voter "types;" or a posterior formed by conditioning on evidence the coalition obtains about voter preferences (e.g., through polling, subterfuge, or other means). This latter indeed seems to be the most likely fashion in which manipulation will proceed in practice. Finally, the standard full knowledge assumption is captured by a point distribution that places probability 1 on the actual vote profile. We sometimes refer to $P$ as a distribution over individual *preferences*, which induces a distribution over profiles by taking the product distribution $P^n$.

The coalition's goal is to cast a collective vote $\mathbf{v}_c$ that maximizes the chance of $d$ winning:

$$\operatorname*{argmax}_{\mathbf{v}_c} \sum_{\mathbf{v}_n:\; r(\mathbf{v}_n,\mathbf{v}_c)=d} P(\mathbf{v}_n).$$

We refer to this $\mathbf{v}_c$ as an *optimal Bayesian manipulation strategy*. For most standard voting rules, this is equivalent to maximizing the *probability of manipulation*, which is the above sum restricted to profiles $\mathbf{v}_n$ such that $r(\mathbf{v}_n) \neq d$.

While we focus on constructive manipulation, our general framework can be applied directly to any reasonable objective on the part of the manipulating coalition. Plausible objectives include: *destructive manipulation*, which attempts to prevent a specific candidate from winning; *safe manipulation*, where a coalitional voter is unsure whether his coalitional colleagues will vote as planned [34, 19]; or *utility maximization*, which attempts to maximize (expected) utility over possible winning candidates. Notice that constructive manipulation can be interpreted as utility maximization with a 0-1 utility for the desired candidate $d$ winning.

### 3.2 ANALYTICAL RESULTS

Our aim is to determine optimal manipulation strategies given any probabilistic beliefs that the coalition might hold, for arbitrary voting rules. Given this general goal, tight analytical results and bounds are infeasible, a point to which we return below. We do provide here two results for optimal manipulation under impartial (and impartial anonymous) culture. Since this style of analysis under partial information is rare, these results suggest the form that further results (e.g., for additional rules and more general distributions) might take. However, as argued elsewhere [32], this preference model is patently unrealistic, so we view these results as being largely of theoretical interest. Indeed, the difficulty in obtaining decent analytical results even for simple voting rules under very stylized distributions strongly argues for a more general computational approach to manipulation optimization that can be applied broadly—an approach we develop in the next section.

We begin with an analysis of the $k$-approval rule. When sincere votes are drawn from the uniform distribution over rankings, each alternative will obtain the same number of approvals in expectation. Intuitively, the coalition should cast its votes so that each approves $d$, and all alternatives apart from $d$ receive the same number of approvals from the coalition (plus/minus 1 if $c(k-1)$ is not divisible by $m-1$): we refer to this as the *balanced strategy*. Indeed, this strategy is optimal:

**Theorem 1.** *The balanced manipulation strategy is optimal for $k$-approval under IC and IAC.*[5]

Things are somewhat more complex for the Borda rule, and we provide results only for the case of three candidates under IC and IAC. Apart from the balanced strategy, we use a *near-balanced strategy*, where the coalition's total approval score for $d$ is $c$, and the scores for the two candidates apart from $d$ differ by at most 2.

**Theorem 2.** *Let $A = \{x, y, d\}$ be a set of three alternatives, assume $c$ is even. Then the either the balanced strategy or the near-balanced strategy is the optimal manipulation strategy for Borda under both IC and IAC. Furthermore, the balanced strategy is optimal if either: (i) $n$ is even and $c + 2$ is divisible by four; or (ii) $n$ is odd and $c$ is divisible by four.*

## 4 A GENERAL OPTIMIZATION FRAMEWORK

Analytical derivation of optimal Bayesian manipulation strategies is difficult; and even for a *fixed* voting rule, it is not viable for the range of beliefs that manipulators might possess about the voter preferences. For this reason, we develop a general optimization framework that can be used to estimate optimal strategies empirically given only the ability to sample vote profiles from the belief distribution. The model will allow direct estimation of the probability of manipulation (and social cost, see below). The model can be adapted to most voting rules, but we focus our development using positional scoring rules for ease of exposition.

---
[5]The nontrivial proofs of the results in this section can be found in the appendix of a longer version of this paper; see: http://www.cs.toronto.edu/~cebly/papers.html.

The main idea is straightforward. Suppose we have a sample of $T$ vote profiles from preference distribution $P$. For each vote profile, a given manipulation will either succeed or not; so we construct an optimization problem, usually in the form of a mixed-integer program (MIP), that constructs the manipulation that succeeds on the greatest number of sampled profiles. If enough samples are used, this approximately maximizes the probability of $d$ winning, or equivalently, the probability of successful manipulation by the coalition. The formulation of the optimization problem—including the means by which one summarizes a sampled vote profile and formulates the objective—depends critically on the voting rule being used. We illustrate the method by formulating the problem for positional scoring rules.

Assume a positional scoring rule using score vector $\alpha$. A sampled vote profile can be summarized by a (summary) score vector $\mathbf{s} = (s_1, \ldots, s_m)$, where $s_i$ is the total score of $a_i$ in that profile; hence we will treat a profile and its score vector interchangeably. Assume $T$ sampled profiles $S = \{\mathbf{s}^1, \ldots, \mathbf{s}^T\}$. A manipulation strategy $\mathbf{v}_c$ can be represented by a PSM $\mathbf{X}$, where $X_{ij}$ denotes the number manipulators who rank candidate $a_i$ in $j$th position. The total score of each candidate for a given profile $\mathbf{s}$ is then $\mathbf{s} + \mathbf{X}\alpha'$.

This strategy representation simplifies the formulation of the optimization problem significantly (by avoiding search over of all possible collections of rankings). Moreover, it is not difficult to recover a set of manipulator votes $\mathbf{v}_c$ that induce any such $\mathbf{X}$, using properties of perfect matchings on $c$-regular bipartite graphs:

**Lemma 3.** *A matrix $\mathbf{X}$ is the PSM for some manipulation strategy $\mathbf{v}_c$ iff $\mathbf{X} \in \mathbb{N}_{\geq 0}^{m \times m}$ and $\mathbf{X}\mathbf{1} = \mathbf{X}'\mathbf{1} = c\mathbf{1}$.*

Our aim then reduces to finding a PSM $\mathbf{X}$ satisfying the above properties such that $\mathbf{X}\alpha'$ maximizes the probability of manipulation. We can recover the optimal manipulation strategy $\mathbf{v}_c^*$ (i.e., a set of $c$ votes) in polynomial time using an algorithm to find $c$ edge-disjoint perfect matchings in a $c$-regular bipartite graph. Specifically, we construct a bipartite graph with candidates forming one set of nodes and "vote positions" forming the second set. We connect these two sets of nodes with a multi-set of edges, with exactly $X_{ij}$ (duplicate) edges connecting candidate $i$ to position $j$. We find a perfect matching in this graph to determine one manipulator vote, remove the corresponding edges, and repeat the process (decreasing each row and column sum by one at each iteration).

We formulate the problem of finding an (approximately) optimal Bayesian manipulation strategy as a MIP which constructs a PSM $\mathbf{X}$ maximizing the number of sampled profiles in $S$ on which $d$ wins. We assume, for ease of exposition only, that $\alpha$ has integral entries. First note that in any optimal strategy, $X_{d1} = c$ and $X_{dj} = 0$ for all $j > 1$, which implies $X_{i1} = 0$ for all $a_i \neq d$. Otherwise we require $X_{ij} \in \{0, \ldots, c\}$ and row and column sum constraints:

$$\sum_{j=2}^{m} X_{ij} = c \quad \forall a_i \neq d, \qquad \sum_{\substack{i=1 \\ i \neq d}}^{m} X_{ij} = c \qquad \forall j > 1.$$

We use variables $I_i^t \in [0,1]$ for all $t \leq T, i \neq d$, where $I_i^t = 1$ iff candidate $a_i$'s total score with manipulators is strictly less than $d$'s total score, constrained as:

$$s_d^t + c\alpha_1 - s_i^t - \sum_{j=2}^{m} X_{ij} \alpha_j \geq$$
$$\alpha_1(n+c)(I_i^t - 1) + 1 \quad \forall t, a_i \neq d. \quad (1)$$

The left-hand side of Eq. (1) is the score difference between $d$ and $a_i$, bounded (strictly) from below by $-\alpha_1(n+c)$. If it is less than 0, then $I_i^t < 1$; otherwise, $I_i^t$ is unconstrained by Eq. (1), and will take value 1 (due to the maximization objective below). Finally, we use variables $I^t \in \{0, 1\}$ to indicate whether $d$ wins under $\mathbf{X}$ on profile $\mathbf{s}^t$, requiring:

$$\sum_{a_i \neq d} I_i^t \geq (m-1)I^t \quad \forall t. \qquad (2)$$

If $d$'s score is less than that of some $a_i$, then the sum in Eq. (2) is smaller than $m-1$, forcing $I^t = 0$ (otherwise the maximization objective will force it to 1). We use the following natural maximization objective:

$$\max_{\mathbf{I}, \mathbf{X}} \sum_{t=1}^{T} I_t. \qquad (3)$$

If $d$ wins in sample $\mathbf{s}$ prior to manipulation (see below), $d$ still wins after manipulator votes are counted, but we do not consider this to be "successful manipulation." Thus, the estimated *probability of manipulation* (distinct from the probability of $d$ winning) is the MIP objective value less the number of profiles in $S$ where $d$ would have won anyway.

The MIP can be simplified greatly. First notice that $d$ cannot win, even with manipulation, in profile $\mathbf{s}^t$ if:

$$s_d^t + c\alpha_1 \leq c\alpha_m + \max_i s_i^t. \qquad (4)$$

Any such profiles—and all corresponding variables and constraints—can be pruned from the MIP. Similarly, we can prune any profile where $d$ wins regardless of the manipulation. This occurs when $d$ wins without manipulator votes or is very close to winning:

$$s_d^t + c\alpha_1 > c\alpha_2 + \max_i s_i^t. \qquad (5)$$

This pruning can greatly reduce the size of the MIP in practice, indeed, in expectation by a factor equal to the probability $P$ that a random profile satisfies condition (4) or (5). The MIP has at most a total of $(T+2)m-2$ constraints, $(m-1)^2+T$ integer variables and $T(m-1)$ continuous variables, where $T$ is the number of non-pruned profiles.

While pruning has a tremendous practical impact, the optimal Bayesian manipulation problem for scoring rules remains NP-hard: this follows from the NP-hardness of Borda manipulation with a known profile [13, 4], and the observation that a single known profile corresponds to a special case of our problem.[6]

The remaining question has to do with sample complexity: in order to have confidence in our estimate, how many samples $T$ should we use? Specifically, if we set a PAC target, obtaining an $\varepsilon$-accurate estimate of the probability of $d$ winning with confidence $1-\delta$, the required number of samples depends on the *VC dimension* $D_\alpha$ of the class of boolean-valued functions over vote profiles (or more generally the corresponding score vectors $\mathbf{s} = (s_1, \ldots, s_m)$):

$$\mathcal{F}_\alpha = \{\mathbf{s} \mapsto \mathbf{1}[d \text{ unique max of } \mathbf{X}\alpha' + \mathbf{s}] \mid \forall \mathbf{X}\}.$$

Using known results [14], on counting $|\mathcal{F}_\alpha|$ one obtains $\sup_\alpha D_\alpha \in O(cm \ln(cm) + c^2)$. Standard sample complexity results then apply directly:

**Proposition 4.** *There exists a constant $C > 0$ such that if $T \geq C(cm \ln(cm) + c^2 + \ln(1/\delta))/\varepsilon^2$ then for any distribution $P$, with probability $1-\delta$ over sample $S$ of size $T$, we have $\hat{q} \leq q^* + \varepsilon$, where $q^*$ is the probability of manipulation of the best strategy, and $\hat{q}$ is the probability of manipulation given the optimal solution to the MIP.*

For specific positional rules, the sample complexity may be smaller. For example, using standard results on compositions of integers, $k$-approval gives rise to a VC dimension of $D_{\text{kappr}} \leq \log_2 \binom{m+ck-1}{ck-1}$, giving the following sample complexity result:

**Proposition 5.** *If $T \geq 256(2\log_2 \binom{m+ck-1}{ck-1} + \ln(4/\delta))/\varepsilon^2$ then for any $P$, with probability $1-\delta$ over sample $S$ of size $T$, we have $\hat{q} \leq q^* + \varepsilon$, where $q^*$ is the probability of manipulation under the best strategy for $k$-approval and $\hat{q}$ is the probability of manipulation given the optimal solution to the MIP for $k$-approval.*

Furthermore, tighter results could be obtained with specific knowledge or constraints on the distribution $P$. Of course, such sample complexity results are conservative, and in practice good predictions can be realized with far fewer samples. Note also that this sample complexity is only indirectly related to the complexity of the MIP, due to the pruning of (typically, a great many) sampled profiles.

---

[6]A partial LP relaxation of the MIP may be valuable in practice: allowing entries of $\mathbf{X}$ to be continuous on $[0,1]$ provides an upper bound on (optimal Bayesian) success probability. Our computational experiments did not require this approximation, but it may be useful for larger problems.

## 5 IMPACT OF MANIPULATION ON SOCIAL WELFARE

As discussed above, characterizing the impact of a manipulating coalition's action solely in terms of its probability of succeeding can sometimes be misleading. This is especially true when one moves away from political domains—where a utility-theoretic interpretation of voting may run afoul of policy, process and fairness considerations—into other settings where voting is used, such as resource allocation, consumer group recommendations, hiring decisions, and team decision making [7]. In such domains, it is often natural to consider the utility that a group member ("voter") derives from the choice or decision that is made for the group as a whole. However, even in "classical" voting situations, most voting protocols are defined using an explicit score, social choice objective, or social welfare function; as such, analyzing the expected loss in this objective due to (optimal) manipulation is a reasonable approach to characterizing the manipulability of different voting rules.

Intuitively, manipulation is more likely to succeed when the desired candidate $d$ is "closer to winning" under a specific voting rule (in the absence of manipulation) than if the candidate is "further from winning." In a (very loose) sense, if candidates that are "closer to winning" are those that are generally ranked more highly by group members, this means that such candidates are generally more desirable. As a consequence, social welfare for alternative $d$ must be close to that of the optimal (non-manipulated) alternative if $d$ has a reasonable chance of winning, which in turn means that the *damage*, or loss in social welfare, caused by manipulation will itself be limited. In this section we formalize this intuition and provide some simple bounds on such damage. We investigate this empirically in the next section.

Assume a voting rule $r$ based on some social welfare measure $SW(a, \mathbf{v})$ over alternatives $a$ and (reported) preference profiles $\mathbf{v}$; hence $r(\mathbf{v}) \in \text{argmax}_a SW(a, \mathbf{v})$. As above, we partition the vote profile: $\mathbf{v} = (\mathbf{v}_n, \mathbf{v}_c)$. We are interested in the loss in social welfare, or *regret*, imposed on the honest voters by a manipulation, so

define this to be:

$$R(\mathbf{v}_n, \mathbf{v}_c) = SW(r(\mathbf{v}_n), \mathbf{v}_n) - SW(r(\mathbf{v}), \mathbf{v}_n).^7 \quad (6)$$

The *expected regret* of a specific manipulation, given distribution $P$ over preference profiles $\mathbf{v}_n$, is then:

$$ER(P, \mathbf{v}_c) = \mathbb{E}_{\mathbf{v}_n \sim P}[R(\mathbf{v}_n, \mathbf{v}_c)]. \quad (7)$$

Notice that any social welfare function $SW$ that determines the quality of a candidate $a$ given the sincere vote profile $\mathbf{v}_n$ can be used in Eq. 6; we need not commit to using $r$'s scoring function itself. However, assessing loss does require the use of *some* measure of societal utility or welfare. If one is concerned only with whether the "true winner" is selected, then probability of manipulation, as discussed above, is the only sensible measure.

We illustrate the our framework by deriving several bounds on loss due to manipulation using positional scoring rules to measure social welfare.[8] We can derive theoretical bounds on expected regret for positional scoring rules. First, notice that expected regret can be bounded for arbitrary distributions:

**Proposition 6.** *Let $r$ be a positional scoring rule with score vector $\alpha$. Then for any distribution $P$, and any optimal manipulation strategy $\mathbf{v}_c$ w.r.t. $P$, we have*

$$ER(P, \mathbf{v}_c) < c[(\alpha_1 - \alpha_m)P(r(\mathbf{v}_n) \neq d \wedge r(\mathbf{v}) = d) \\ + (\alpha_2 - \alpha_m)P(r(\mathbf{v}_n) \neq r(\mathbf{v}) \wedge r(\mathbf{v}) \neq d)]. \quad (8)$$

Intuitively, this follows by considering the maximum increase in score the manipulating coalition can cause for $d$ relative to an alternative $a$ that would have won without the manipulators.

*Proof of Prop. 6.* Consider any $\mathbf{v}_n$. Clearly, $d$ is ranked first by all votes in $\mathbf{v}_c$. Case 1: if $d$ wins in $\mathbf{v}_n$ then $d$ also wins on $\mathbf{v}$. Case 2: if $a_i$ wins in $\mathbf{v}_n$ but $d$ wins on $\mathbf{v}$ then $SW(a_i) + \alpha_m c \leq SW(a_i) + SW(a_i, \mathbf{v}_c) < SW(d) + \alpha_1 c$ implying $R(\mathbf{v}_n, \mathbf{v}_c) <$ $c(\alpha_1 - \alpha_m)$. Case 3: if $a_i$ wins in $\mathbf{v}_n$ and $a_j \neq a_i$ wins in $\mathbf{v}_c$ then $SW(a_i) + \alpha_m c \leq SW(a_i) + SW(a_i, \mathbf{v}_c) \leq SW(a_j) + \alpha_2 c$ implying $R(\mathbf{v}_n, \mathbf{v}_c) \leq c(\alpha_2 - \alpha_m)$. Case 4: if $r(\mathbf{v}_n) = r(\mathbf{v}_c)$ then regret is zero. Summing 2 and 3 gives the upper bound on expected regret. □

---
[7] We assume a voting rule and welfare measure that can accept variable numbers of voters, as is typical.

[8] One reason to consider positional scoring rules like Borda in analyzing impact on social welfare is the tight connection between scoring rules and social welfare maximization in its narrow sense (i.e., sum of individual utilities). In models where we desire to maximize sum of utilities relative to some underlying utility profile, voting is a relatively simple and low-cost way (i.e., with minimal communication) of eliciting partial preference information from voters. Analysis of the *distortion* of utilities induced by restricting voters to expressing ordinal rankings shows that, with carefully crafted positional rules, one can come close to maximizing (this form of) social welfare [37, 29, 6]. Borda scoring, in particular, seems especially robust in this sense.

The proposition applies when $P$ reflects full knowledge of $\mathbf{v}_n$ as well; and while this $P$-dependent bound will be crude for some $P$, it is in fact tight in the worst-case (which includes full knowledge distributions):

**Proposition 7.** *Suppose $\alpha_1 + \cdots + \alpha_m = M$, $m-1$ divides $c$ and $n - c \geq 0$ is even. Then*

$$\sup_{n,c,\alpha} \sup_P ER(P, \mathbf{v}_c) = cM. \quad (9)$$

*Proof.* The upper bound on the LHS follows from the RHS of Eq. 8 since it is at most $c(\alpha_1 - \alpha_m) \leq cM$. For the lower bound on the LHS, let $P$ be a point mass on $\{\mathbf{v}_n\}$: in $\mathbf{v}_n$, the first $c$ votes rank $a_1$ first and the remaining alternatives $a_2, \ldots, a_{m-1}, p$ in such a way that the number of times any is ranked $i$-th ($i \geq 2$) is $c/(m-1)$. Of the remaining $n - c$ votes, half rank $a_1$ first and $d$ second, and half do the opposite (the remaining candidates are ranked in any manner). Thus $SW(a_1) - SW(d) = (\alpha_1 - \frac{\alpha_2 + \cdots + \alpha_m}{m-1})c$. Let $\alpha_2 = \delta + \xi$ and $\alpha_i = \delta$, for some $\delta, \xi > 0$, for all $i \geq 3$. One optimal strategy $\mathbf{v}_c$ is to always place $d$ first and $a_1$ last, resulting in a score difference of $c(\alpha_1 - \alpha_m) = c(\alpha_1 - \delta)$ which is strictly larger than the above social welfare difference of $c(\alpha_1 - \delta - \xi/(m-1))$ within $\mathbf{v}_n$. Hence $\mathbf{v}_c$ causes $d$ to win, inducing regret of $c(M - (m-1)\delta - \xi m/(m-1))$, which can be made arbitrarily close to $cM$ using a small enough $\delta, \xi$. □

We can obtain a tighter bound than that offered by Prop. 6 if, for a given $P$ and $r$, we know the optimal manipulation strategy. For instance, exploiting Thm. 1 we obtain:

**Proposition 8.** *Consider the k-approval rule, where $SW(a, \mathbf{v}_n)$ is the approval score of $a$. Let $P$ be impartial culture. Then*

$$ER(\mathbf{v}_n, BAL) \leq \left[\left\lceil \frac{ck}{m} \right\rceil - 1\right] \cdot [P(r(\mathbf{v}_n) \neq d \wedge r(\mathbf{v}) = d) \\ + P(r(\mathbf{v}_n) \neq r(\mathbf{v}) \wedge r(\mathbf{v}) \neq d)]. \quad (10)$$

*Proof.* Consider any $\mathbf{v}_n$. We use a case analysis similar to that in Prop. 6. Case 1 applies directly. For case 3, $a_i$ must have received one more veto vote than $a_j$ from the manipulators, and thus $SW(a_i) - SW(a_j) \leq 1$. For case 2, BAL implies that $SW(a_i) - \lceil \frac{ck}{m} \rceil + 1 \leq SW(p)$ ($a_i$ might have received $\lfloor ck/m \rfloor$ veto votes, but the bound holds in any case). Thus $SW(a_i) - SW(d) \leq \lceil \frac{ck}{m} \rceil - 1$. Case 4 also applies directly. Summing cases 2 and 3 gives the required inequality. □

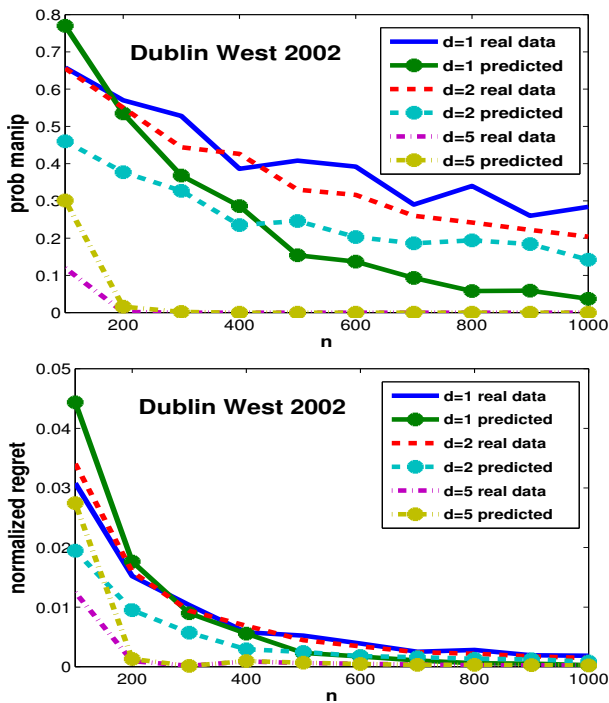

**Fig. 1:** Probability of manipulation and expected normalized regret for Irish election data.

We can also exploit our empirical optimization framework to assess the expected regret. While deriving optimal manipulation strategies is analytically intractable in general, we can exploit the ability to empirically determine (approximately) optimal Bayesian strategies to great effect. Given a collection of samples, we can—in the same MIP used to compute the optimal strategy—measure expected regret empirically.[9] Sample complexity results related to those above can be derived (we omit details). Notice the ability to accurately estimate the behavior of the manipulators is critical to being able to estimate expected regret.

## 6 EMPIRICAL EVALUATION

We experiment with several common preference distributions, as well as real voting data, to test the effectiveness of our approach. We are primarily interested in: (a) determining the computational effectiveness of our empirical framework for computing optimal manipulation strategies when manipulators have incomplete knowledge; and (b) measuring both the probability of manipulation and expected regret caused by manipulation in some prototypical scenarios. We focus on Mallows models, and mixtures of Mallows models, because of their flexibility and ability to represent a wide class of preferences (including impartial culture).

However, it should be clear that our framework is not limited to such models. Space precludes a systematic investigation of multiple voting rules, so we focus here on the Borda rule.

Our first set of experiments uses 2002 Irish electoral data from the Dublin West constituency, with 9 candidates and 29,989 ballots of top-$t$ form, of which 3800 are complete rankings.[10] We learn a Mallows mixture with three components (using the techniques of [21]) using a subsample of 3000 random (not necessarily complete) ballots: this represents a prior that manipulators might develop with intense surveying. We fix $c = 10$ manipulators and vary the number of sincere voters $n \in \{100, 200, \ldots, 1000\}$. The MIP is solved using 500 random profiles sampled from the learned Mallows mixture, which provides the manipulators with an approximately optimal strategy, as well as predicted probability of success and expected regret. We test the predictions by simulating the manipulators' strategy on the real data set, drawing 1000 random profiles (of the appropriate size $n$) from the set of 3800 full rankings to estimate true probability of manipulation and expected regret (normalized by $SW(r(\mathbf{v}_n), \mathbf{v}_n)$). Since manipulability varies greatly with the "expected rank" of a candidate, we show results for the candidates whose *expected* ranks in the learned model are first, second, and fifth (below this, probability of manipulation is extremely small).

Fig. 1 shows that the probability of manipulation is in fact quite high when $d$ is the first- or second-ranked candidate (in expectation), but is much smaller when $d$ is the fifth-ranked. Not surprisingly, probabilities gradually drop as $n$ grows. The predicted probabilities based on the learned model are reasonably accurate, but of course have some error due to imperfect model fit. Despite the high probability of manipulation, the second plot shows that expected regret (normalized to show percentage loss) is in fact extremely small. Indeed, maximal (average) loss in social welfare is just over 3%, when $d$ is candidate 2 and $n = 100$, which means means nearly 10% of the voters are manipulators. Expected regret drops rapidly with increasing $n$. Notice that success probability and expected regret are greatest when the manipulators' desired candidate has expected rank 1 or 2: while the odds of 1 winning are higher than 2, 1 is also more likely to win without manipulator intervention.

Our second set of experiments use Mallows models over six alternatives with different variance $\phi$ (recall that with $\phi = 1$, Mallows is exactly IC). The reference ranking is $\sigma = 123456$ (i.e., alternative 1 is most preferred, 2 next most, etc.). We fix $c = 10$ and vary $n$ from 100 to 1000 as above. Results in Fig. 2 show manipulation probability and expected regret as we vary $\phi$ and consider desired alternatives 1, 2 and 3.[11] While manipulation probability is high for these near-top alternatives (when $n$ is small), expected regret (normalized in percentage terms) is negligible, with a maximum of 4%, and then only when the distribution is close to impartial culture ($\phi = 0.8$) and $n = 100$ (nearly 10% manipulators). As above, when manipulators want $d = 2$, expected regret is highest. Of some interest is the connection to both theoretical and empirical work that shows phase transitions often occur when the number of manipulators is roughly the square root of the number of sincere voters: any less makes manipulation very unlikely, while any more makes manipulation likely. While most of this work analyzes complete information settings, our results above show that *with realistic preference distributions*—even with restricted knowledge on the part of manipulators—the probability of manipulation is sometimes quite significant with far fewer manipulators than suggested by past work. Despite this, expected regret remains relatively small.

---
[9]Pruning of samples must be less aggressive when estimating regret, since damage may be caused by manipulation even in profiles where $d$ cannot win.

[10]See www.dublincountyreturningofficer.com.

| $\phi$ | $d$ | $n$:100 | 200 | 300 | 400 | 500 | 600 |
|---|---|---|---|---|---|---|---|
| .6 | 1 | .03 | .00 | 0 | 0 | 0 | 0 |
| | 2 | .46 | .06 | .00 | 0 | 0 | 0 |
| .8 | 1 | .19 | .09 | .05 | .03 | .01 | .01 |
| | 2 | .68 | .41 | .25 | .13 | .08 | .05 |
| | 3 | .41 | .07 | .01 | 0 | 0 | 0 |
| 1 | * | .46 | .34 | .26 | .21 | .17 | .17 |

| $\phi$ | $d$ | $n$:100 | 200 | 300 | 400 | 500 |
|---|---|---|---|---|---|---|
| .6 | 1 | 5.3E-4 | 1.7E-5 | 0 | 0 | 0 |
| | 2 | 3.2E-2 | 2.1E-3 | 7.1E-5 | 0 | 0 |
| .8 | 1 | 5.5E-3 | 1.7E-3 | 8.0E-4 | 2.2E-4 | 7.5E-5 |
| | 2 | 4.0E-2 | 1.4E-2 | 6.0E-3 | 2.5E-3 | 1.3E-3 |
| | 3 | 9.6E-3 | 7.6E-4 | 7.4E-5 | 1.4E-5 | 1.2E-5 |
| 1 | * | 1.9E-2 | 7.1E-3 | 3.6E-3 | 2.5E-3 | 1.5E-3 |

**Fig**. 2: Prob. of manipulation (top) and expected normalized regret (bottom), Mallows models.

| sec. | $n$:100 | 200 | 300 | 400 | 500 | 600 |
|---|---|---|---|---|---|---|
| avg | 28.67 | 0.35 | 0.20 | 0.18 | 0.06 | 0.07 |
| max | 205.64 | 2.56 | 1.17 | 1.16 | 0.12 | 0.15 |

| sec. | $d$ | $\phi$:0.5 | 0.6 | 0.7 | 0.8 | 0.9 | 1 |
|---|---|---|---|---|---|---|---|
| avg | 2 | 0.02 | 0.04 | 0.02 | 0.02 | .05 | 2.44 |
| | 4 | 0.02 | 0.03 | 0.02 | 0.02 | 0.14 | |
| max | 2 | 0.03 | 0.15 | 0.03 | 0.07 | 0.14 | 30.22 |
| | 4 | 0.03 | 0.08 | 0.05 | 0.04 | 1.16 | |

**Fig**. 3: MIP solution times for Dublin (top) and Mallows (bottom) on an 8-core 2.66GHz/core machine.

Fig. 3 shows the average and maximum running times of the MIP required to compute the optimal manipulation for the problems described above. As can be seen, even with a large number of sampled profiles, the MIP can be solved quickly across a range of problem sizes and distributions.

---

[11]Under IC (i.e., when $\phi = 1$), alternatives are probabilistically indistinguishable, so we show one row only.

## 7 CONCLUDING REMARKS

Our primary contribution is an empirical framework for the computation of optimal manipulation strategies when manipulators have incomplete information about voter preferences. This is an important methodology for the analysis of the manipulability of voting rules in realistic circumstances, without the need to restrict the analysis to specific voting rules or priors. Our experiments indicate that our algorithms are quite tractable. Furthermore, our results suggest that manipulation may not be as serious a problem as is commonly believed when realistic informational models are used, or when the quality of the outcome, rather than societal justice, is the main objective. Our empirical results, which exploit several innovations introduced in this paper, demonstrate this in the case of Borda voting; but our approach is easily adapted to different types of manipulation under different scoring rules, and can be applied to any "utility-based" voting rule with appropriate formulation of the optimization problem. Thus, our approach provides a compelling framework for the comparison of voting rules.

One nonstandard aspect of our approach is the use of the score under a voting rule as a proxy for social welfare. A similar regret-based approach is taken in preference elicitation [22]; and it is implicit in work on approximating voting rules [28, 5], which assumes that approximating the score also gives an approximation to the desirability of an alternative. One strong argument in favor of this view is that scores of certain voting rules, such as Borda, are provably good proxies for utilitarian social welfare when utilities are drawn from specific distributions [37]. That said, our framework can be used, in principle, to analyze any measure of impact or loss.

Our work suggests a number of interesting future directions. Of course, we must study additional voting rules, social welfare measures, and manipulator objectives within our framework to further demonstrate its viability. While our results suggest that incomplete knowledge limits the ability of a manipulating coalition to impact the results of an election, our framework can also be used to directly study the relationship between the "amount of information" (e.g., using entropy or equivalent sample size metrics) and the probability of manipulation. Finally, interesting computational questions arise within our approach: e.g., deriving tighter complexity results for optimal manipulation given a *collection* of vote profiles; or deriving simpler classes of manipulation policies (e.g., uncertainty-sensitive variants of the balanced manipulation strategy) that can be more readily optimized by a manipulating coalition.